\documentclass[aps,prl,twocolumn,groupedaddress]{revtex4}
\usepackage{graphicx}
\usepackage{amsmath}
\usepackage[sort&compress]{natbib}
\usepackage{hyperref}

\newcommand{\pr}{\theta}     
\newcommand{\p}{\varphi}		
\newcommand{\pscale}{a}
\newcommand{\Mob}{\mathbf{M}}         
\newcommand{\Mobc}{M}         
\newcommand{\kstiff}{k_{\infty}}         
\newcommand{\PD}{p}         
\newcommand{\Uhess}{\mathbf{e}}
\newcommand{\Uhessc}{e}
\newcommand{\Sep}{\mathbf{U}}
\newcommand{\Sepc}{U}
         
\newcommand{\NIDcoeff}{\mathbf{B}}         
\newcommand{\NIDcorr}{c}         
\newcommand{\qsaddle}{\hat{q}}         
\newcommand{\dU}{\Delta V}         
         
\newcommand{\curv}{\gamma}         
\newcommand{\kT}{k_{\mathrm{B}}T}
\newcommand{\EQ}[1]{Eq.~(\ref{eq:#1})}

\newcommand{\FIG}[1]{Fig.~\ref{fig:#1}}

\begin{document}

\title{Optimal flexibility for conformational transitions in macromolecules}

\author{Richard A. Neher$^{1}$} 
\author{Wolfram M\"obius$^{1}$}
\author{Erwin Frey$^{1}$}
\author{Ulrich Gerland$^{1,2}$}

\affiliation{$^{1}$Arnold Sommerfeld Center for Theoretical Physics (ASC) and Center for Nanoscience (CeNS),\\ LMU M\"unchen, Theresienstra{\ss}e 37, 80333 M\"unchen, Germany\\ $^{2}$Institute for Theoretical Physics, University of Cologne, Germany}

\date{\today}

\begin{abstract}
Conformational transitions in macromolecular complexes often involve the reorientation of lever-like structures. Using a simple theoretical model, we show that the rate of such transitions is drastically enhanced if the lever is bendable, e.g.~at a localized ``hinge''. Surprisingly, the transition is fastest with an intermediate flexibility of the hinge. In this intermediate regime, the transition rate is also least sensitive to the amount of ``cargo'' attached to the lever arm, which could be exploited by molecular motors. To explain this effect, we generalize the Kramers-Langer theory for multi-dimensional barrier crossing to configuration dependent mobility matrices. 
\end{abstract}

\maketitle

Many biological functions depend on transitions in the global conformation of macromolecules, and the associated kinetic rates can be under strong evolutionary pressure. For instance, the directed motion of molecular motors is based on power strokes \cite{Howard2001}, protein binding to DNA can require DNA bending \cite{Sugimura_PNAS_2006} or spontaneous partial unwrapping of DNA from histones \cite{Li:05, Moebius_PRL_06}, and the functioning of some ribozymes depends on global transitions in the tertiary structure \cite{XiaoweiZhuang05242002}. These and other examples display two generic features: 
(i)~A long segment within the molecule or complex is turned during the transition, e.g.~an RNA stem in a ribozyme, the DNA as it unwraps from histones or bends upon protein binding, or the lever arm of a molecular motor relative to the attached head. 
(ii)~The segment has a certain bending flexibility. 
Here, we use a minimal physical model to study the coupled dynamics of the transition and the bending fluctuations. 

\begin{figure}[b]
\includegraphics[width=0.8\columnwidth]{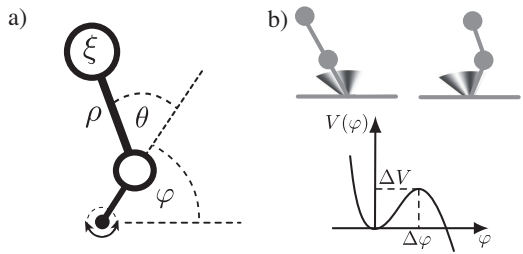}
\caption{\label{fig:model}
Schematic illustration of the `Two-Segment Lever' (TSL) model for conformational transitions. 
(a) The two segments of lengths $1$ and $\rho$ are connected by a hinge and attached to the origin. The viscous drag acts on the ends of the segments as indicated by the beads. 
(b) Schematic illustration of the barrier crossing processes. The external meta-stable potential $V(\p)$ is indicated by shading (top; dark corresponds to high energy) and is also sketched below. 
}
\end{figure}

Our model, illustrated in \FIG{model}, demonstrates explicitly how even a small bending flexibility can drastically accelerate the transition. Furthermore, if the flexibility arises through a localized ``hinge'', 
e.g.~in the protein structure of some molecular motors \cite{Terrak_PNAS_05, Jeney_ChemPhysChem_04} or an interior loop in an RNA stem, we find that the transition rate is maximal at an intermediate hinge stiffness. Thus, in situations where rapid transition rates are crucial, molecular evolution could tune a hinge stiffness to the optimal value. 
We find that an intermediate stiffness is optimal also from the perspective of robustness, since it renders the transition rate least sensitive to changes in the drag on the lever arm, incurred e.g.~by different cargos transported by a molecular motor. 

Our finding of an optimal rate is reminiscent of a phenomenon known as resonant activation \cite{Doering_PRL_92, Reimann_PRL_95}, where a transition rate displays a peak as a function of the characteristic timescale of fluctuations in the potential barrier. However, we will see that the peak in our system has a different origin: a trade-off between the accelerating effect of the bending fluctuations and a decreasing average mobility of the reaction coordinate. The standard Kramers-Langer theory \cite{Langer_APNY_69} for multi-dimensional transition processes is not sufficient to capture this trade-off. A generalization of the theory to the case of configuration-dependent mobility matrices turns out to be essential to understand the peak at intermediate stiffness.

{\it Model.---}
We model the conformational transition as a thermally activated change in the attachment angle $\p$ of a macromolecular lever, see \FIG{model}. The lever has two segments connected by a hinge with stiffness $\epsilon$, which renders the lever preferentially straight, but allows thermal fluctuations in the bending angle $\pr$. The energy function $V(\p,\pr)$ of this `Two-Segment Lever' (TSL) is 
\begin{equation}
\label{eq:potential} 
\frac{V(\p,\pr)}{\kT}=\epsilon(1-\cos \pr)-\left[\frac{(\pscale\p)^{3}}{3}-\frac{b(\pscale\p)^{2}}{2}\right] \;,
\end{equation}
where $k_{B}T$ is the thermal energy unit. 
The hinge, described by the first term, serves not only as a simple model for a protein or RNA hinge, but also as a zeroth-order approximation to a more continuously distributed flexibility; see below. The second term is the potential  on the attachment angle $\p$, which produces a metastable minimum at $(\p,\pr)=(0,0)$. The thermally-assisted escape from this minimum passes through the transition state at $(\p,\pr)=(b/\pscale, 0)$ with a barrier height $\dU=b^3\kT/6$ \footnote{With the potential (\ref{eq:potential}) this transition is irreversible, however all of our conclusions equally apply to reversible transitions in a double-well potential.}.

In the present context, inertial forces are negligible, i.e.~it is sufficient to consider the stochastic dynamics of the TSL in the overdamped limit. We localize the friction forces to the ends of the two segments, as indicated by the beads in \FIG{model}(a). The length of the first segment defines our length unit and $\rho$ denotes the relative length of the second segment. Similarly, we choose our time unit such that the friction coefficient of the first bead is unity, and denote the coefficient of the second bead by $\xi$. To describe the Brownian dynamics of the TSL, we derive the Fokker-Planck equation for the time-evolution of the configurational probability density $p(\p,\pr,t)$. In general, the derivation of the correct dynamic equations can be a nontrivial task for stochastic systems with constraints  \cite{Helfand_JChemPhys_79, vanKampen_AJP_84}. For instance, implementing fixed segment lengths through the limit of stiff springs, leads to Fokker-Planck equations with equilibrium distributions that depend on the way in which the limit is taken \cite{vanKampen_AJP_84}. However, for our overdamped system, we can avoid this problem by imposing the desired equilibrium distribution, i.e.~the Boltzmann distribution $p=\exp(-V/\kT)$, which together with the well-defined deterministic equations of motion uniquely determines the Fokker-Planck equation for the TSL. 

The deterministic equations of motion take the form $\dot{q}_k=-\Mob_{kl}\,\partial V/\partial q_l$ with the coordinates $(q_1, q_2)=(\p,\pr)$ and a mobility matrix $\Mob$. We obtain $\Mob$ with a standard Lagrange procedure: Given linear friction, $\Mob$ is the inverse of the friction matrix, which in turn is the Hessian matrix of the dissipation function \cite{Goldstein}. This yields 
\begin{equation}
\label{eq:mobility} 
\Mob=\frac{1}{1+\xi \sin^2 \pr}\left(\begin{array}{cc}
1& \frac{\rho+\cos \pr}{\rho}\\[0.1cm]
\frac{\rho+\cos \pr}{\rho}&\frac{\rho+2\cos \pr}{\rho}+ \frac{1+\xi}{\xi\rho^{2}}
\end{array}\right)\;.
\end{equation}
The Fokker-Planck equation then follows from the continuity equation $\partial_t \PD(\{q_i\},t) = -\partial_k j_k (\{q_i\},t)$ together with \begin{equation}
\label{eq:fluxdensity} 
j_k(\{q_i\},t) 
= -\Mobc_{kl}\left[\frac{\partial V}{\partial q_l}+\kT\frac{\partial}{\partial q_l} \right] \PD(\{q_i\},t)
\end{equation}
as the probability flux density. Our analytical analysis below is based directly on Eqs.~(\ref{eq:mobility}) and (\ref{eq:fluxdensity}), while we perform all Brownian dynamics simulations with a set of equivalent stochastic differential equations \cite{Gardiner_04}.

\begin{figure}
\includegraphics[width=0.8\columnwidth]{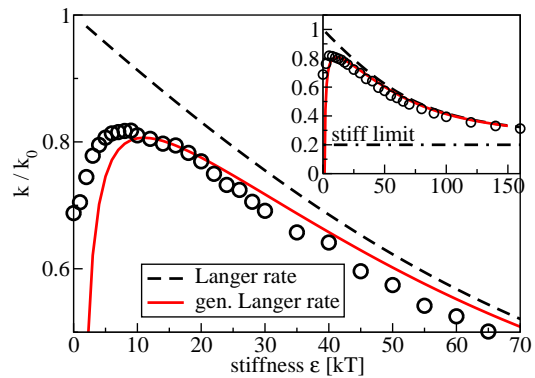}
\caption{\label{fig:rate_vs_stiffness} 
Simulation data of the barrier crossing rate normalized by $k_0$ display a prominent peak at finite stiffness (circles, each obtained from 20000 simulation runs initialized at the metastable minimum).  
The conventional Langer theory fails to describe the non-monotonicity of the rate and overestimates 
the rate at small $\epsilon$. The generalized Langer theory captures the non-monotonicity
of the rate and describes the simulations data accurately; parameters see main text. 
}
\end{figure}

{\it Transition rate.---}
To explore the phenomenology of the TSL, we performed simulations to determine its average dwell time $\tau$ in the metastable state, for a range of hinge stiffnesses $\epsilon$. The rate for the conformational transition is related to the dwell time by $k(\epsilon)=1/\tau(\epsilon)$. \FIG{rate_vs_stiffness} shows $k(\epsilon)$ (circles) for a barrier $\dU\!=\!12\,\kT$, a distance $\Delta\p\!=\!0.4$ to the transition state, and $\xi\!=\!\rho\!=\!1$ (data for different parameter values behaves qualitatively similar, as long as the process is reaction-limited, i.e. $\dU$ is sufficiently large that $\tau$ is much longer than the time for the TSL to freely diffuse over an angle $\Delta\p$). We observe a significant flexibility-induced enhancement of the transition rate over a broad range of stiffnesses, compared to the dynamics in the stiff limit ($\epsilon\!\to\!\infty$), see inset. Note that the enhancement persists even at relatively large $\epsilon$, where typical thermal bending fluctuations $\delta\p\sim\sqrt{\epsilon}$ are significantly smaller than $\Delta\p$. Surprisingly, the acceleration is strongest at an intermediate stiffness ($\epsilon\approx 10$). This observation suggests that the stiffness of molecular hinges could be used, by evolution or in synthetic constructs, to tune and optimize reaction rates. 
 
\begin{figure}
\includegraphics[width=0.8\columnwidth]{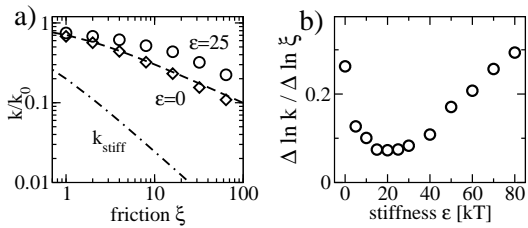}
\caption{\label{fig:rate_vs_friction} 
The sensitivity of the rate to the friction coefficient $\xi$ is minimal at intermediate stiffness.
(a) Simulation results at $\epsilon=0$ and $\epsilon=25$ as well as the theoretical estimates of the rate at $\epsilon=0$ and in the stiff limit.
(b) The derivative of $\ln k$ with respect to $\ln \xi$ evaluated at $\xi=1$, i.e.~the slope of the curves in a), is minimal in an intermediate
stiffness range. 
}
\end{figure}

When the friction coefficient $\xi$ of the outer bead is increased, the rate of the conformational transition decreases; see \FIG{rate_vs_friction}a. This decrease is most dramatic in the stiff limit (dash-dotted line). In the flexible limit (diamonds) the decrease is less pronounced. Notably, the rate appears least sensitive to the viscous drag on the outer bead at intermediate $\epsilon$ (circles). Indeed, \FIG{rate_vs_friction}b shows that the $\epsilon$-dependence of this sensitivity (measured as the slope of the curves in \FIG{rate_vs_friction}a at $\xi\!=\!1$) has a pronounced minimum at $\epsilon\approx 20$. Hence, intermediate hinge stiffnesses in the TSL lead to maximal robustness, which is an important design constraint for many biomolecular mechanisms in the cellular context. For instance, as molecular motors transport various cargos along one-dimensional filaments, it may be advantageous to render their speed insensitive to the cargo size, e.g. to avoid ``traffic jams''.

In the remainder of this letter, we seek a theoretical understanding of the above phenomenology. First, it is instructive to consider simple bounds on the transition rate. An upper bound is obtained by completely eliminating the outer bead. The Kramers rate \cite{Haenggi_RevModPhys_90} for the remaining 1D escape process, $k_0=(\pscale^2b/2\pi)\,e^{-\dU/\kT}$, is used in Figs.~\ref{fig:rate_vs_stiffness} and \ref{fig:rate_vs_friction} to normalize the transition rates. At the optimal stiffness, the transition rate in \FIG{rate_vs_stiffness} comes within $20\,\%$ of this upper bound. 
An obvious lower bound is the stiff limit: For $\epsilon\!\to\!\infty$, the second segment increases the rotational friction by a factor $\zeta=1+(1+\rho)^2\xi$, so that the 1D Kramers rate becomes $\kstiff=k_0/\zeta$, as shown by the dash-dotted line in  \FIG{rate_vs_stiffness} and \FIG{rate_vs_friction}a. 
However, to understand how the dynamics of the bending fluctuations affects the transition rate, we must consider the full 2D dynamics of the TSL. The multi-dimensional generalization of Kramers theory is Langer's formula for the escape rate over a saddle in a potential landscape \cite{Langer_APNY_69}, 
\begin{equation}
\label{eq:Langer}
k_{\rm Langer}=\frac{\lambda}{2\pi} \times \sqrt{\frac{\det \Uhess^{({\rm w})}}{|\det \Uhess^{({\rm s})}|}}\, \exp\left(-\frac{\dU}{\kT}\right)\;.
\end{equation}
Here, $\Uhess^{({\rm w})}$ and $\Uhess^{({\rm s})}$ denote the Hessian matrix of the potential energy, $\partial^{2} V/\partial q_{k}\partial q_{l}$, evaluated at the well bottom and the saddle point, respectively, whereas $\lambda$ is the unique negative eigenvalue of the product of the mobility matrix $\Mob$ and $\Uhess^{({\rm s})}$. 
\EQ{Langer} can be made plausible in simple terms:
Given a quasi-equilibrium in the metastable state, the second factor represents the probability of being in the transition region, i.e. the region within $\sim \kT$ of the saddle. 
The escape rate is then given by this probability multiplied by the rate $\lambda$ at which the system relaxes out of the transition state, analogous to Michaelis-Menten reaction kinetics. 

For our potential (\ref{eq:potential}), the determinants in (\ref{eq:Langer}) cancel. The eigenvalue can be determined analytically (the dashed line in \FIG{rate_vs_stiffness} shows the resulting $k_{\rm Langer}$), but for the present purpose it is more instructive to consider the expansions for large and small stiffness. In the stiff limit, the natural small parameter is the stiffness ratio $\curv/\epsilon$, where $\gamma=a^{2}b$ is the absolute curvature or ``stiffness'' of the external potential at the transition state. The expansion yields $k_{\rm Langer}/\kstiff=1+(\rho^2\xi/\zeta)\,\curv/\epsilon+\mathcal{O}(\curv^{2}/\epsilon^{2})$. As expected, the rate approaches $\kstiff$, but the stiff limit is attained only when the bending fluctuations $\sim\sqrt{\epsilon}$ are small compared to the width of the barrier $\sim\sqrt{\curv}$. In the opposite limit, $\epsilon\!\ll\!\curv$, the rate is given by 
$k_{\rm Langer}/k_0=1-\left(1+\rho^{-1}\right)^{2}\epsilon/\curv+\mathcal{O}(\epsilon^{2}/\curv)$.
Since the linear term is negative, Langer theory predicts that the transition rate peaks at zero stiffness, with a peak value equal to the Kramers rate $k_0$ for the lever without the second segment. This prediction is clearly at variance with the simulation results. It is interesting to note, however, that the slope of the linear decay is independent of $\xi$. This is consistent with our observation that the transition rate is insensitive to $\xi$ in the intermediate stiffness regime. Indeed, \FIG{rate_vs_stiffness} shows that Langer theory (dashed line) describes the simulation data (circles) reasonably well for intermediate and large hinge stiffness.

\begin{figure}
\includegraphics[width=0.8\columnwidth]{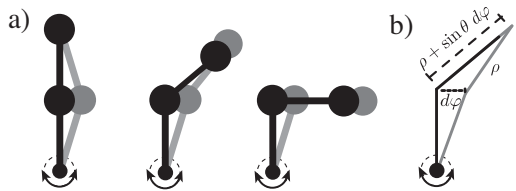}
\caption{\label{fig:mobility} 
The friction opposing rotation of the attachment angle $\p$ depends on the bending angle $\pr$,
since the outer bead is moved by different amounts in different configurations.
For an infinitesimal displacement $d\p$, the displacement of the outer bead is $\sin\pr\;d\p$.
The projection of the resulting friction force onto the direction of motion adds another factor $\sin\pr$,
yielding a friction coefficient for $\p$ of $1+\xi\sin^{2}\pr$.   
}
\end{figure}

To understand the origin of the peak at intermediate stiffness, it is useful to consider the flexible limit ($\epsilon=0$). In this limit, the transition state is degenerate in $\pr$, and it seems plausible to estimate the transition rate by using a $\pr$-averaged mobility for the reaction coordinate $\p$,  
\begin{equation}
\label{eq:floppy}
k(\epsilon=0)\approx k_0\int_{-\pi}^{\pi}\frac{d\pr}{2\pi} \Mobc_{11}(\pr)=\frac{k_0}{\sqrt{1+\xi}}\;.
\end{equation}
This estimate agrees well with the simulation data, see the dashed line in \FIG{rate_vs_friction}a, indicating that the configuration-dependent mobility (\ref{eq:mobility}) plays an important role for the transition rate. In contrast, the conventional Langer theory assumes the mobility matrix to be constant 
in the relevant region near the transition state. \FIG{mobility} illustrates why the mobility $\Mobc_{11}$ of the coordinate $\p$ is affected by the bending angle $\pr$ and gives a graphical construction for $\Mobc_{11}$.

{\it Generalized Langer theory.---}
To account for the mobility effect identified above, we must generalize the Langer theory to configuration-dependent mobility matrices. The special case where the mobility varies only along the reaction coordinate has already been studied in \cite{Gavish_PRL_80}, however the main effect in our case is due to the variation in the transverse direction. 
In the following, we outline the derivation of the central result, while all details will be presented elsewhere. Near the saddle, the mobility matrix takes the form 
$\Mobc_{ij}(\{q_i\})=\Mobc_{ij}^{({\rm s})}+\frac{1}{2}A^{kl}_{ij}\qsaddle_l\qsaddle_k$, where $\qsaddle_i$ are deviations from the saddle and $A^{kl}_{ij}$ denotes the tensor of second derivatives of the mobility matrix (we assume that the first derivatives of $\Mob$ vanish at the saddle, which is the case for the TSL). The escape rate is given by the probability flux out of the metastable well, divided by the population inside the well. To calculate the flux, we construct a steady state solution to the Fokker-Planck equation in the vicinity of the saddle, as described in \cite{Haenggi_RevModPhys_90} for the conventional Langer theory. We use the Ansatz 
$\PD(\{q_i\})=\frac{1}{2}\PD_{eq}(\{q_i\})\,\mathrm{erfc}(u)$, where $\PD_{eq}(\{q_i\})=Z^{-1}e^{-V(\{q_i\})/\kT}$ and $\mathrm{erfc}(u)$ is the complementary error function with argument $u=\Sepc_k \qsaddle_k$. Inserting the Ansatz into the Fokker-Planck equation yields an equation for the vector $\Sep$, 
\begin{equation}
\label{eq:eigenvectoreq}
\Sepc_i(-\Mobc_{ij}\Uhessc_{jk}^{({\rm s})}+B_{ik})-\Sepc_i \Mobc_{ij} \Sepc_j \; \Sepc_k =0 \;,
\end{equation}
where $B_{ik}=\kT\sum_{n} A_{ni}^{nk}$. $B_{ik}\qsaddle_k$ is the noise induced drift, which is absent in 
the conventional Langer theory. 
Ignoring higher order terms, this equation determines $\Sep$ to be the left eigenvector of 
$-\Mob^{({\rm s})}\Uhess^{({\rm s})}+\NIDcoeff$ to the unique positive eigenvalue
$\lambda$, and requires $\Sep$ to be normalized such that $\Sepc_i M^{({\rm s})}_{ij}\Sepc_j=\lambda$. The directions of the left and right eigenvectors of $-\Mob^{({\rm s})}\Uhess^{({\rm s})}+\NIDcoeff$ have a physical interpretation: $\Sep$ is perpendicular to the stochastic separatrix, while the corresponding right eigenvector points in the direction of the diffusive flux at the saddle \cite{Berezhkovskii_JChemPhys_05}.

From $\PD(\{q_i\})$, the flux density is determined by (\ref{eq:fluxdensity}) and the total flux is obtained by integrating the flux density over a plane containing the saddle; a convenient choice is the plane $u=0$. Evaluation of the integral is particularly simple in a coordinate system, where the first coordinate is parallel to $\Sep$, and the remaining coordinates are chosen such that $\Uhess^{({\rm s})}$ is diagonal in this subspace, $\Uhessc_{ij}^{({\rm s})}=\mu_i \delta_{ij}$ for $i,j>1$. In this coordinate system, the generalized Langer rate takes the simple form
\begin{equation}
\label{eq:genLangerrate}
k=\frac{\lambda}{2\pi} \frac{1+\frac{1}{2 \Mobc_{11}}\sum_{l>1} \frac{A^{ll}_{11}}{\mu_l}}{\sqrt{1-\NIDcorr}} \times \sqrt{\frac{\det \Uhess^{({\rm w})}}{|\det \Uhess^{({\rm s})}|}} \, e^{-\frac{\dU}{\kT}}\;,
\end{equation}
where $\NIDcorr=\Sepc_i\Uhess^{-1}_{ij}\Sepc_j+1=B_{1i}\Uhessc^{-1}_{i1}/\Mobc_{11}^{({\rm s})}$ and $\Uhess^{-1}$ denotes the inverse matrix of $\Uhess^{({\rm s})}$. 
\EQ{genLangerrate} contains three corrections to (\ref{eq:Langer}), all of which vanish when $\Mob(\{q_i\})$ is constant: The most important one is given by $\sum_{l>1} A^{ll}_{11}/\mu_l$, which changes the mobility $\Mobc_{11}$ in the direction of $\Sep$ to an effective mobility that is averaged over the separatrix with respect to the Boltzmann distribution. In addition, there are two corrections incurred by the noise-induced drift: the factor $\sqrt{1-\NIDcorr}$ and a change due to the fact that $\lambda$ is now the eigenvalue to $\Mob^{({\rm s})}\Uhess^{({\rm s})}-\NIDcoeff$
instead of $\Mob^{({\rm s})}\Uhess^{({\rm s})}$. 

The solid line in \FIG{rate_vs_stiffness} shows the application of the generalized Langer formula to the TSL. We observe that it captures the peak in the transition rate and thus the essential phenomenology of the TSL. Obviously, the evaluation of the Gaussian integral that leads to \EQ{genLangerrate} is only meaningful, if the harmonic approximation of the mobility matrix is reasonable within the relevant 
saddle point region. This integral diverges as the saddle point degenerates, which explains the behavior for $\epsilon\to 0$. At high $\xi$, the very anisotropic friction can also render Langer theory invalid \cite{Drozdov_JChemPhys_96, Berezhkovskii_ChemPhys_91}.

{\it Conclusion.---}
We have introduced the ``Two-Segment Lever'' as a simple model for a class of conformational transitions in biomolecules. The model clearly demonstrates how flexibility can enhance the rate of a conformational transition. This remains true, if the hinge in the TSL is replaced by a more continuous bendability. Interestingly, a discrete hinge has a stiffness regime, where the rate is large and robust against cargo variation, which raises the question, whether these effects are exploited by evolution, for example in the design of molecular motors. To understand these effects theoretically, we derived a generalized Langer theory that takes into account configuration dependent mobility matrices. We hope that this theory will find applications also in other fields. 

We thank the German Excellence Initiative for financial support via the program NIM. RN and UG are grateful for the hospitality of the CTBP at UCSD, where part of this work was done, and for financial support by the CeNS in Munich and the DFG.

\end{document}